\documentclass{iopart}

\usepackage[dvips]{graphicx}
\usepackage{bm}

\begin{document}

\paper{Entanglement Evolution in the Presence of Decoherence}

\author{Jin~Wang$^{1,2}$, Herman~Batelaan$^1$, Jeremy~Podany$^1$, and Anthony~F.~Starace$^1$}

\address{$^1$Department of Physics and Astronomy, The University of
Nebraska, Lincoln, NE 68588-0111\\ $^2$Department of Physics, The
University of Tennessee, Chattanooga, TN 37403-2598}

\begin{abstract}
The entanglement of two qubits,  each defined as an effective
two-level, spin 1/2 system, is investigated for the case that the
qubits interact via a Heisenberg XY interaction and are subject to
decoherence due to population relaxation and thermal effects. For
zero temperature, the time dependent concurrence is studied
analytically and numerically for some typical initial states,
including a separable (unentangled) initial state. An analytical
formula for non-zero steady state concurrence is found for any
initial state, and optimal parameter values for maximizing steady
state concurrence are given. The steady state concurrence is found
analytically to remain non-zero for low, finite temperatures. We
also identify the contributions of global and local coherence to the
steady state entanglement.
\end{abstract}


\newcommand{\beq}{\begin{equation}}
\newcommand{\eeq}{\end{equation}}
\newcommand{\bqa}{\begin{eqnarray}}
\newcommand{\eqa}{\end{eqnarray}}
\newcommand{\nn}{\nonumber}
\newcommand{\nl}[1]{\nn \\ && {#1}\,}
\newcommand{\erf}[1]{Eq.~(\ref{#1})}
\newcommand{\rf}[1]{(\ref{#1})}
\newcommand{\dg}{^\dagger}
\newcommand{\rt}[1]{\sqrt{#1}\,}
\newcommand{\smallfrac}[2]{\mbox{$\frac{#1}{#2}$}}
\newcommand{\half}{\smallfrac{1}{2}}
\newcommand{\brra}[1]{\langle{#1}|}
\newcommand{\ket}[1]{|{#1}\rangle}
\newcommand{\ip}[2]{\langle{#1}|{#2}\rangle}
\newcommand{\sch}{Schr\"odinger }
\newcommand{\schs}{Schr\"odinger's }
\newcommand{\hei}{Heisenberg }
\newcommand{\heis}{Heisenberg's }
\newcommand{\bl}{{\bigl(}}
\newcommand{\brr}{{\bigr)}}
\newcommand{\ito}{It\^o }
\newcommand{\str}{Stratonovich }
\newcommand{\dbd}[1]{\frac{\partial}{\partial {#1}}}
\newcommand{\sq}[1]{\left[ {#1} \right]}
\newcommand{\cu}[1]{\left\{ {#1} \right\}}
\newcommand{\ro}[1]{\left( {#1} \right)}
\newcommand{\an}[1]{\left\langle{#1}\right\rangle}
\newcommand{\implies}{\Longrightarrow}



\section{\label{Intro}Introduction}

Entanglement is a property of correlated quantum systems that cannot
be accounted for classically. Entangled states of distinct (possibly
interacting) quantum systems, which are those that cannot be
factorized into product states of the subsystems, are of fundamental
interest in quantum mechanics. The production of pairwise entangled
states  is an essential requirement for the operation of the quantum
gates that make quantum information and quantum computation possible
\cite{Bouwmeester}. Considerable attention has been devoted to
interacting Heisenberg spin systems
\cite{Arnesen,Wang,gerard,Korepin}, which serve as a model for
various solid state \cite{Loss1, Loss2, Imamoglu} or NMR
\cite{Ernst, Nielson} quantum computation schemes and for simulating
magnetic phenomena in condensed matter systems using atoms in
optical lattices~\cite{SorensenMolmer, Duan}.  Indeed, general
Hamiltonians that include Heisenberg spin-spin interactions have
been proposed as ``generic''~\cite{Shepelyansky} or
``ideal''~\cite{Makhlin} model Hamiltonians for quantum computation
systems.  A key question for entangled quantum states is the effect
of decoherence due to the environment (see, e.g.,~\cite{Bloch,
Zurek, Albrecht,Das Sarma, Dodd} and references therein), which is
not only a fundamental issue for quantum computation
devices~\cite{Lloyd, Knight} but also for the relation between
quantum and classical physics~\cite{Zurek, Braun}.  Although there
have been many investigations of decoherence in recent years,
careful investigation of well-understood model systems continue to
produce surprises that add to fundamental understanding.  For
example, Yu and Eberly~\cite{Eberly} have recently shown that the
entanglement of a pair of non-interacting qubits in the presence of
spontaneous decay of the upper states may decohere in a finite time
instead of exponentially.

In this paper we examine decoherence due to both population
relaxation and thermal effects for an entangled (and interacting)
two qubit system.  The Hamiltonian for our two-qubit system has the
form of the well-known Heisenberg XY model for two interacting spins
in the presence of an external magnetic field, where the effective
magnetic field is defined by the energy separation of the two-level
system that we associate with each (spin 1/2) qubit.  As noted
above, this form of Hamiltonian is very common in models for quantum
computing~\cite{SorensenMolmer, Duan, Shepelyansky,Makhlin}.   Our
analysis of decoherence complements that of Ref.~\cite{Eberly} by
examining a system in which the qubits interact.  For our two-qubit
model system at zero temperature, we find that for any initial
state, including the common one in which the two qubits are
initially unentangled, the system reaches a steady state of pairwise
entanglement in spite of population relaxation. The extent of
steady-state entanglement  is sensitive to both the spatial
anisotropy of the interaction between qubits and to the energy level
separation of the two levels associated with each qubit.  To the
extent that these two parameters can be varied in some particular
physical realization of our model system, the magnitude of steady
state entanglement may thus be controlled.  We analyze both
analytically and numerically the time-dependent evolution of the
entanglement (as measured quantitatively by the
concurrence~\cite{Bennett, Wootters}) of our model two-qubit system
for some typical initial states: a pure, separable initial state; a
pure, entangled initial state; and a mixed initial state. We also
obtain an analytic formula for the steady state concurrence that
shows its dependence on both the system parameters and the
decoherence rate and that enables us to specify optimal values for
these parameters to achieve the maximum possible concurrence. In a
separate section, we consider the case of finite temperature and
present an analytic formula for the concurrence, which remains
non-zero over a finite range of low temperatures. In our concluding
section, we discuss some implications of these results.

\section{\label{Hamiltonian}Two-Qubit Hamiltonian}

We note that the Hamiltonian of a Heisenberg chain of $N$ spin
$\frac{1}{2}$ particles with nearest-neighbor interactions
is~\cite{Korepin}:

\beq
H=\sum_{n=1}^N(J_xS_{n}^{x}S_{n+1}^{x}+J_yS_{n}^{y}S_{n+1}^{y}+J_zS_{n}^{z}S_{n+1}^{z})
\eeq where $S_{n}^{\alpha}=\frac{1}{2}\sigma_{\alpha}^{n}
(\alpha=x,y,z)$ are the local spin $\frac{1}{2}$ operators at site
$n$, the $\sigma_{\alpha}^{n}$ operators are the Pauli matrices at
site $n$,  the periodic boundary condition $S_{N+1}=S_1$ applies,
and $\hbar=1$. For arbitrary $J_{\alpha}$'s, the Heisenberg chain is
often called the $XYZ$ model. The chain is said to be
antiferromagnetic for $J_{\alpha}>0$ and ferromagnetic for
$J_{\alpha}<0$. The $XY (J_z=0)$ and the Heisenberg-Ising
$(J_y=J_z=0)$ interactions have been analyzed for nuclear spin
systems \cite{Ernst}, in particular for nuclear magnetic resonance
approaches to quantum computation (see, e.g., Section 7.7 of
Ref.~\cite{Nielson}).

The Hamiltonian $H$ for an anisotropic two-qubit Heisenberg $XY$
system in an (effective) external magnetic field $\omega$ along the
z-axis  is: \beq\label{Hamilton}
H=\omega(S_{1}^{z}+S_{2}^{z})+J(S_{1}^{+}S_{2}^{-}+S_{1}^{-}S_{2}^{+})+\Delta(S_{1}^{+}S_{2}^{+}+S_{1}^{-}S_{2}^{-})
\eeq where $J=(J_x+J_y)/2$, $\Delta=(J_x-J_y)/2$, and
$S^{\pm}=S^{x}{\pm}{i}S^{y}$ are the spin raising and lowering
operators. The first term in the Hamiltonian describes the energy of
the spins in the effective external magnetic field. This effective
field is defined by the energy levels of our qubits:  we assume that
each of our two qubits represents an identical two level system
whose two energies are defined by $\pm \omega/2$. The spin
interaction Hamiltonian, described by the second and the third terms
in Eq.~({\ref{Hamilton}}), produces the coherence of the two qubits
that is necessary for their entanglement in the presence of
decoherence. As shown below, the third term, whose magnitude is
proportional to the parameter $\Delta$, which describes the spatial
anisotropy of the spin-spin interaction, is essential for the
production of steady state entanglement.

\section{\label{Zero}Time Evolution of the Concurrence at Zero Temperature}

The time evolution of the system for zero temperature, $T=0$, is
given by the following master equation (see, e.g., \cite{Carmichael,
Gardiner} and Section 8.4.1 of \cite{ Nielson}):

\beq \label{Lindblad}\dot{\rho}=-i\left[H,\rho\right] + \gamma{\cal
D}\left[S_{1}^{-}\right]\rho+\gamma{\cal
D}\left[S_{2}^{-}\right]\rho.\eeq Here $\rho$ is the density matrix,
which in the presence of population relaxation represents the mixed
state of the system. The Lindblad super operator $\cal D$
{~\cite{lindblad}} is defined by ${\cal D}[A]B \equiv ABA\dg -
\{A\dg A,B\}/2$,
 which describes the population relaxation of the upper state of each qubit due to the environment;
$\gamma$ is the phenomenological rate of population relaxation,
which for simplicity is assumed to be the same for each of the two
qubits (i.e., we assume each qubit has the same interaction with the
environment).  As discussed below, the assumption of a single decay
rate, $\gamma$, requires us to place restrictions on the magnitude
of the coupling between qubits.

Entanglement is increasingly regarded as a physical resource of a
quantum information system (see, e.g., Section 12.5 of
Ref.~\cite{Nielson}) and many measures for quantifying entanglement
have been developed (see, e.g., \cite{Bennett, Wootters, Vedral,
Horodecki, Rains, Preskill}). Since decoherence processes cause the
system state to become mixed, we use the measure of entanglement
termed concurrence  \cite{Bennett, Wootters}.   For a system
described by the density matrix $\rho$, the concurrence $C$  is
\beq \label{C} C={\rm
max}\left(\sqrt{\lambda_1}-\sqrt{\lambda_2}-\sqrt{\lambda_3}-\sqrt{\lambda_4},0\right),
\eeq where $\lambda_1$, $\lambda_2$, $\lambda_3$, and $\lambda_4$
are the  eigenvalues (with $\lambda_1$ the
largest one) of the ``spin-flipped'' density operator $R$, which is
defined by \beq R =\rho
\left(\sigma_{y}\otimes\sigma_{y}\right)\rho^{*}
\left(\sigma_{y}\otimes\sigma_{y}\right), \eeq where $\rho^{*}$
denotes the complex conjugate of ${\rho}$  and ${\sigma}_{y}$ is the
usual Pauli matrix.  $C$ ranges in magnitude from $0$ to $1$;
nonzero $C$ denotes an entangled state.

The basis states $\ket{\psi_{i}}$ for our two-qubit system are the
separable product states of the individual qubits: \bqa
\label{basis4}
\ket{\psi_{1}}&=&\ket{e}_{1}\otimes\ket{e}_{2},\nn \\
\ket{\psi_{2}}&=&\ket{e}_{1}\otimes\ket{g}_{2},\nn \\
\ket{\psi_{3}}&=&\ket{g}_{1}\otimes\ket{e}_{2},\nn \\
\ket{\psi_{4}}&=&\ket{g}_{1}\otimes\ket{g}_{2}. \eqa  In general, a
two-qubit system is represented by a  density matrix having sixteen
non-zero elements. For our Hamiltonian, however, the density matrix
can be represented as the sum of two submatrices that evolve
independently of one another,  \bqa \label{densitymatrix}
\rho=\left(\begin{array}{cccc}
\rho_{11}& 0  & 0 &\rho_{14}\\
0& \rho_{22}  & \rho_{23} &0 \\
0& \rho_{32}  & \rho_{33} &0 \\
\rho_{41}& 0  & 0 &\rho_{44} \\
\end{array}\right)+
\left(\begin{array}{cccc}
0 & \rho_{12}  & \rho_{13} & 0 \\
\rho_{21}& 0  & 0 & \rho_{24} \\
\rho_{31}& 0  & 0 & \rho_{34} \\
0 & \rho_{42}  & \rho_{43} & 0 \\
\end{array}\right),
\eqa i.e.,  in solving  Eq.~({\ref{Lindblad}}) for $\rho(t)$  the
forms of each of the two submatrices in Eq.~(\ref{densitymatrix})
are preserved.  (Note that the second matrix on the right hand side of Eq.~(\ref{densitymatrix}) does not have the form of a density matrix.)  Each of the examples in this paper has an initial density matrix whose form is that of the first matrix on the right of  Eq.~(\ref{densitymatrix}).  This limitation is not very restrictive, as unentangled, entangled, and mixed states can all be described. 
Furthermore, for a state having a density matrix of the
form of the first matrix on the right of Eq.~(\ref{densitymatrix}),
the concurrence has the following analytic form: \bqa\label{CR} C=\max{\{0, C_1,
C_2\}},\eqa  where
 \bqa\label{C1C2}
C_1&=&2(|\rho_{41}|-\sqrt{\rho_{33}\rho_{22}})\nn\\
C_2&=&2(|\rho_{32}|-\sqrt{\rho_{44}\rho_{11}}).
 \eqa  

The solutions of the master equation in Eq.~({\ref{Lindblad}})
simplify by transforming from the product state basis
$\ket{\psi_{i}}$ in Eq.~({\ref{basis4}}) to the basis of eigenstates
$\ket{\Phi_{\alpha}}$ of the Hamiltonian in Eq.~({\ref{Hamilton}}),
\bqa \label{basis5}
\ket{\Phi_{1}}&=&N^+(\ket{gg} + \frac{\Delta}{\Omega-\omega}\ket{ee}),\nn \\
\ket{\Phi_{2}}&=&\frac{1}{\sqrt{2}}(\ket{eg}+\ket{ge}),\nn \\
\ket{\Phi_{3}}&=&\frac{1}{\sqrt{2}}(\ket{ge}-\ket{eg}),\nn \\
\ket{\Phi_{4}}&=&N^-(\ket{gg}-\frac{\Delta}{\Omega+\omega}\ket{ee}),\\
\label{omega}
\Omega &=& \sqrt{\omega^2+\Delta^2},\\
N^{\pm}&=&(\Omega \mp \omega)/\sqrt{\Delta^2 + (\Omega \mp
\omega)^2}. \eqa
After transformation, the solutions for each element
$\bar{\rho}_{{\alpha}{\alpha'}}$ of the density matrix (where
$\bar{\rho}$ denotes $\rho$ in the eigenstate basis) can  be found
analytically.  For the interesting special case that both qubits are
initially in their ground states (i.e., the system is initially in
state $\ket{\psi_{4}}$  in Eq.~(\ref{basis4})), the analytic
solutions for $\bar{\rho}_{{\alpha}{\alpha'}}(t)$ are:

\begin{eqnarray}\label{element}
\label{p11}
\bar{\rho}_{11}(t)&=&\frac{1}{2\Omega{\alpha}}\Big[-\omega\alpha+2\Omega{\Delta^2}e^{-2\gamma{t}}\nn\\
&&+\Omega(\alpha-2\Delta^2)+2e^{-\gamma{t}}\Delta^2\gamma{\sin{[2\Omega{t}]}}\Big]\\
\label{p22}
\bar{\rho}_{22}(t)&=&\frac{\Delta^2}{\Omega{\alpha}}\Big[\Omega-{\Omega}e^{-2\gamma{t}}-e^{-\gamma{t}}\gamma{\sin{[2\Omega{t}]}}\Big]\\
\label{p33}
\bar{\rho}_{33}(t)&=&\bar{\rho}_{22}(t)\\
\label{p44}
\bar{\rho}_{44}(t) &=&1-\bar{\rho}_{11}(t)-\bar{\rho}_{22}(t)-\bar{\rho}_{33}(t)\\
\label{p14}
\bar{\rho}_{14}(t)&=&\frac{\Delta}{4i\Omega^2+2\Omega\gamma}\Big[2i\Omega
e^{-\gamma t}\cos{[2\Omega t]}\nn\\
&&+2\Omega e^{-\gamma t}\sin{[2\Omega t]}+\gamma\Big]\\
\label{p41} \bar{\rho}_{41}(t)&=&\bar{\rho}_{14}^*(t)
\end{eqnarray}
where all other matrix elements are zero and where
\begin{eqnarray}
\label{alpha} \alpha&=&4\Omega^2+\gamma^2.
\end{eqnarray}

From Eqs.~(\ref{p11}-\ref{p41}) it is evident that both the
off-diagonal (coherence) matrix elements $\bar{\rho}_{14}$ and
$\bar{\rho}_{41}$ in Eqs.~(\ref{p14}-\ref{p41}) and the diagonal
(population) matrix elements $\bar{\rho}_{{\alpha}{\alpha}}$ in
Eqs.~(\ref{p11}-\ref{p44}) have terms that oscillate with frequency
$2\Omega$. Note that the coherence matrix elements $\bar{\rho}_{14}$
and $\bar{\rho}_{41}$ are non-zero only when the spin-spin
interactions are anisotropic (i.e., $\Delta \neq 0$); also, the
value of $\Omega$ is sensitive to this anisotropy (cf.
Eq.~(\ref{omega})).  From Eqs.~(\ref{p11}-\ref{p41}) it can be seen
that the coherence matrix elements have terms that decay at the rate
$\gamma$ while the population matrix elements also have terms that
decay at the rate $2\gamma$. Analytic solutions similar to
Eqs.~(\ref{p11}-\ref{p41}) can be given for some other initial
states.

The assumption of a single decay rate, $\gamma$, in the master
equation ({\ref{Lindblad}}) requires some discussion.  Owing to the
interaction between qubits described by the Hamiltonian
({\ref{Hamilton}}), the two-qubit energy level structure is altered
from that describing non-interacting, identical qubits.
Nevertheless, the assumption of a single decay rate, $\gamma$, is
reasonable provided the interaction does not significantly alter the
effective energy level separations, or, more precisely, provided the
rotating wave approximation remains valid~\cite{ZollerPC} (see,
e.g., pp. 160-161 of Ref.~\cite{Gardiner}).  The eigenstates in
Eq.~(\ref{basis5}) have the following eigenenergies~\cite{gerard}:
the Bell states have eigenvalues $\pm J$ while the other eigenstates
have eigenvalues $\pm \Omega$.  Thus if we restrict the magnitudes
of the coupling parameter $J$ and the anisotropy parameter $\Delta$
to values such that,
\begin{eqnarray}
 &&|J|/\omega \le{0.1}\label{restriction1}\\
 &&(\Omega - \omega)/\omega \le{0.1} \label{restriction2},
\end{eqnarray}
we shall ensure that the energy level separations of the interacting
qubit system do not change by more than~10\% from that of the
non-interacting qubit system.  Except where it is explicitly
mentioned otherwise, all examples given below have parameter values
for which the above inequalities are satisfied.

\begin{figure}
\begin{center}
\includegraphics[width=8.0cm]{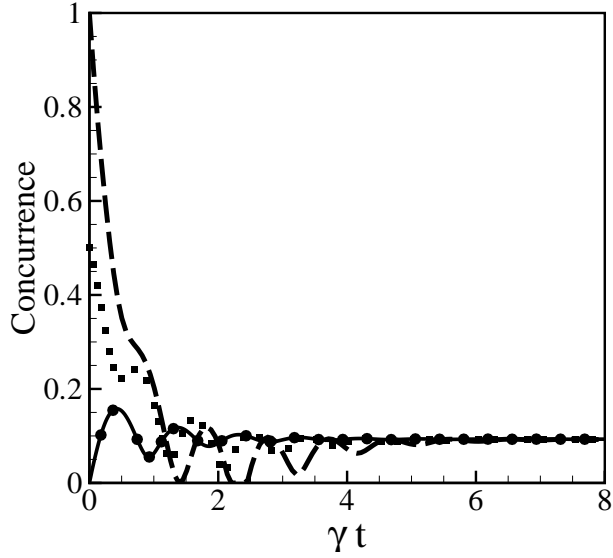}
\end{center}
\caption{\label{fig1}Plot of $T=0$ concurrence vs. scaled time,
$\gamma t$, for three different initial states: (1) An initially
unentangled state, $\ket{\Psi}=\ket{gg}$ (solid line); an initially
entangled state, the Bell state $\ket{\Psi}= \frac{1}{\sqrt{2}}(
\ket{gg}-\ket{ee})$ (dashed line); and (3) an initially mixed state,
defined as an equal mixture of  $\ket{gg}$ and  $\frac{1}{\sqrt{2}}(
\ket{eg}+\ket{ge})$ (solid squares). The system parameters are:
$\gamma=0.3$, $\omega=1.0$, $J=0.1$, and $\Delta=0.1$.}
\end{figure}

\begin{figure}
\begin{center}
\includegraphics[width=8.0cm]{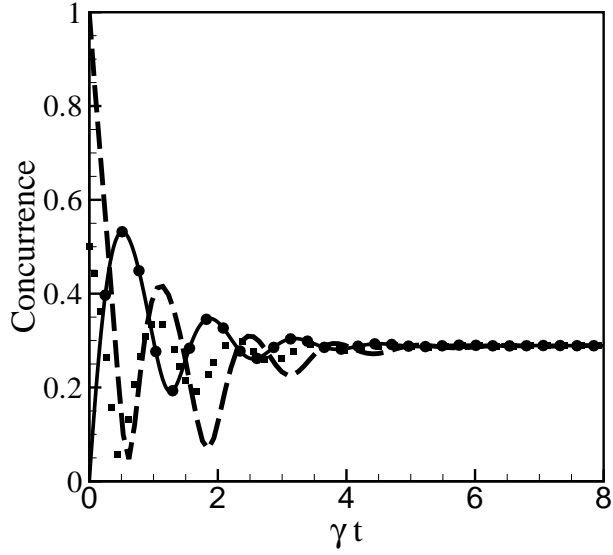}
\end{center}
\caption{\label{fig2}Plot of $T=0$ concurrence vs. scaled time,
$\gamma t$, for the same three initial states as in Fig.~\ref{fig1},
but for the following different system parameters:  $\gamma=0.458$,
$\omega=1.0$, $J=0.1$, and $\Delta=0.458$.}
\end{figure}

Perhaps surprisingly, the decoherence due to population relaxation
does not prevent the creation of a steady state level of
entanglement, regardless of the initial state of the system. This is
demonstrated in Figs.~{\ref{fig1}} and~\ref{fig2}, which show the
time evolution of the concurrence (cf. Eq.~(\ref{C})) for three
different initial states: (1)  An unentangled, separable state,
$\ket{\psi_{4}}$ (cf. Eq.~(\ref{basis4})); (2) a completely
entangled state, the Bell state $\ket{\Psi}=
\frac{1}{\sqrt{2}}(\ket{gg}-\ket{ee})$; and (3) a mixed state,
defined as an equal mixture of $\ket{\psi_{4}}$ and the Bell state
$\ket{\Phi_{2}}$.  In Fig.~\ref{fig1}  we consider the case that $J
= \Delta = \omega/10$, which implies that $J_y = 0$ and which thus
corresponds to the ``generic'' quantum computation model Hamiltonian
of Ref.~\cite{Shepelyansky}.  In Fig.~\ref{fig2}  we consider the
case that $J = \omega/10$ and that $\Delta = 0.458 \omega$, which
corresponds to a general case in which $J_x$ and $J_y$ have opposite
signs, which may possibly be achieved for an optical lattice
system~\cite{SorensenMolmer, Duan}.  For each of the three initial
states considered, the corresponding curves  in Figs.~{\ref{fig1}}
and~\ref{fig2}  give the numerical results for the concurrence
defined by Eq.~(\ref{C}), after solving Eq.~(\ref{Lindblad})
numerically for the density matrix in the separable representation
(cf. Eq.~(\ref{basis4})).  Since each  initial state has a density
matrix of the form of the first matrix on the right of
Eq.~(\ref{densitymatrix}), the concurrence for each of these states
is given also by Eqs.~(\ref{CR}-\ref{C1C2}). (Note that
discontinuities in the time derivatives of $C(t)$ for the dashed
curve in Fig.~\ref{fig1} in the range $2.0~{\leq}~t ~{\leq} ~2.5$
stem from the definition in Eq.~(\ref{C}); all density matrix
elements are smooth functions of $t$.) The solid circles on the
curves for the initial state $\ket{gg}$ in Figs.~{\ref{fig1}}
and~\ref{fig2}  give the concurrence  obtained from the analytic
Eqs.~(\ref{CR}-\ref{C1C2}) (after transforming the analytic
expressions in Eqs.~(\ref{p11}) - (\ref{p41}) for this state's
density matrix $\bar{\rho}_{{\alpha}{\alpha'}}$ to $\rho_{ij}$).
These analytic results coincide with those obtained by direct
numerical solution of Eq.~(\ref{Lindblad}).

Despite the presence of decoherence, the results in Figs.~\ref{fig1}
and~\ref{fig2} show that the concurrence reaches the same steady
state value (after some oscillatory behavior) for a given set of
system parameters, regardless of the initial state of the system.
(This is true even for initial states having non-zero matrix
elements belonging to the second matrix on the right of
Eq.~(\ref{densitymatrix}); for our system, such matrix elements
vanish in the steady state.)   Clearly the Heisenberg spin-spin
interaction in Eq.~(\ref{Hamilton}) serves to maintain an entangled
state despite the presence of decoherence in Eq.~(\ref{Lindblad}).
We find a steady state concurrence of 0.09309 for the system
parameter values chosen in Fig.~\ref{fig1} and a steady state
concurrence of 0.28916 for the system parameter values chosen in
Fig.~\ref{fig2}.

The analytic expressions for the $T=0$ steady state values of
$\rho_{ij}(t)$ are as follows:
\begin{eqnarray}
\label{p11a}
\rho_{11}&=& \rho_{22}=\rho_{33}=\frac{\Delta^2}{\alpha}\\
\label{p44a}
\rho_{44}&=& 1 - \frac{3\Delta^2}{\alpha}\\
\label{p14a}
\rho_{14}&=&\frac{-2\omega\Delta-i\Delta\gamma}{\alpha}\\
\label{p41a} \rho_{41}&=&\rho_{14}^*
\end{eqnarray}
The corresponding $T=0$ steady state concurrence is found to be:
 \beq \label{steadycon}
C_{steady}=\frac{2\sqrt{\Delta^2(4\omega^2+\gamma^2)}-2\Delta^2}{\alpha}
\eeq This result stems from $C_1$ in Eq.~({\ref{C1C2}}) calculated
for the separable basis density matrix $\rho_{ij}$  in
Eqs.~(\ref{p11a}-\ref{p41a}). The steady-state concurrence is seen
to depend on the system parameters $\omega$,  $\Delta$, and $\gamma$
(but not on $J$). Also, $\gamma$ serves as a scale factor, i.e.,
$C_{steady}$ depends only on the scaled variables,
$\bar{\omega}=\omega/\gamma$ and $\bar{\Delta}=\Delta/\gamma$. These
parameters may be varied in order to maximize $C_{steady}$.   The
function $C_{steady}(\bar{\omega},\bar{\Delta})$ is shown in
Fig.~{\ref{fig3}; one sees that the surface has a ridge along which
it takes its maximum value. The coordinates of the ridge and the
value of $C_{steady}$ on the ridge may be determined analytically.
For fixed $\omega$, $C_{steady}$ (cf. Eq.~(\ref{steadycon})) has its
maximum at the following value of $\Delta$:
\begin{eqnarray}
\label{Deltamax} \Delta_{max}&=& \frac{\sqrt{4 \omega^2 +
\gamma^2}}{(1+\sqrt{5})}.
\end{eqnarray}
The solid line in the $\bar{\omega}$-$\bar{\Delta}$ plane of
Fig.~{\ref{fig3}} represents the locus of points
$\bar{\Delta}_{max}(\bar{\omega})$ given by Eq.~(\ref{Deltamax})
(upon division by $\gamma$). Substitution of $\Delta_{max}$ into
Eq.~(\ref{steadycon}) gives the parameter-independent maximum value
of the concurrence, represented by the solid line in
Fig.~{\ref{fig3}} along the ridge of $C_{steady}$:
\begin{eqnarray}
\label{Csteadybmax} C_{steady}(\Delta_{max})&=&
(1+\sqrt{5})^{-1}=0.309.
\end{eqnarray}
Eq.~(\ref{steadycon}) shows that in order to have a positive value
of $C_{steady}$, one must have $4\omega^2+\gamma^2 \geq\Delta^2$.
Note finally that Eq.~(\ref{Csteadybmax}) was derived from
Eq.~(\ref{steadycon}) without taking into account the restrictions
on the parameter values imposed by the conditions in
Eqs.~(\ref{restriction1}) and (\ref{restriction2}) that are
necessitated by our assumption of a single decay rate, $\gamma$.
Nevertheless, one sees for the example plotted in Fig.~{\ref{fig2}}
that there do exist values of the parameters that satisfy
Eqs.~(\ref{restriction1}) and (\ref{restriction2}) for which one
obtains a steady state level of concurrence that is close to the
global maximum value given by Eq.~(\ref{Csteadybmax}) (and shown by
the solid line in Fig.~{\ref{fig3}}).

\begin{figure}
\begin{center}
\includegraphics[width=8.0cm]{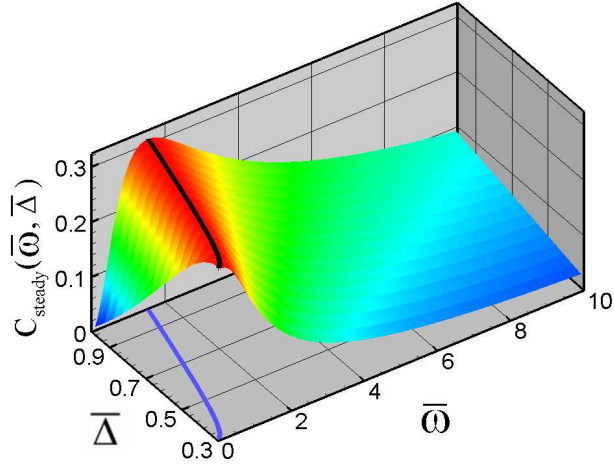}
\end{center}
\caption{\label{fig3}Plot of the $T=0$ steady state concurrence
(cf. Eq.~(\ref{steadycon})) as a function of the scaled energy
$\bar{\omega}$ and the scaled anisotropy parameter $\bar{\Delta}$
(ranging from $0.309$ to $1$), where $\bar{\omega}=\omega/\gamma$
and $\bar{\Delta}=\Delta/\gamma$. The solid lines locate the maximum
value of concurrence (cf. Eq.~(\ref{Csteadybmax})); see text for
discussion.}
\end{figure}

\section{\label{Finite}Temperature Dependence of the Steady State Concurrence}

It is of interest to examine how the steady state entanglement
obtained for zero temperature in the prior section changes when the
temperature is finite. For simplicity, we assume that each qubit
interacts with the same thermal bath.  It is known that the
equilibrium entanglement must vanish at some finite temperature
{\cite{Fine}}.  In order to examine the effect of thermal
decoherence on the entanglement for our system we consider the
following master equation {\cite{Carmichael2, Mintert}:

\bqa \label{Lindblad2}\dot{\rho}&=&-i\left[H,\rho\right] +
\gamma{(\bar{n}+1)}{\cal
D}\left[S_{1}^{-}\right]\rho+\gamma{(\bar{n}+1)}{\cal
D}\left[S_{2}^{-}\right]\rho \nn\\&&+\gamma\bar{n}{\cal
D}\left[S_{1}^{+}\right]\rho+\gamma\bar{n}{\cal
D}\left[S_{2}^{+}\right]\rho, \eqa where $\bar{n}$, the average
excitation of the thermal bath,  parametrizes the temperature. Note
that $\bar{n}$ is zero at zero temperature, whereupon one observes
that Eq.~(\ref{Lindblad2}) reduces to Eq.~(\ref{Lindblad}); also,
$\bar{n}$ becomes infinite as the temperature becomes infinite.  The
master equation (\ref{Lindblad2}) may be solved to obtain the
following analytic expressions for the steady state density matrix
of our system:

\bqa\label{densitymatrixT}
\rho_{11}&=&\frac{\bar{n}^2(4\bar{\omega}^2+(1+2\bar{n})^2)+\bar{\Delta}^2(1+2\bar{n})^2}{(1+2\bar{n})^2(4\bar{\omega}^2+(1+2\bar{n})^2+4\bar{\Delta}^2)}\nn\\
\rho_{22}&=&\rho_{33}=\frac{1}{4}[1-\frac{4\bar{\omega}^2+(1+2\bar{n})^2}{(1+2\bar{n})^2(4\bar{\omega}^2+(1+2\bar{n})^2+4\bar{\Delta}^2)}]\nn\\
\rho_{44}&=&\frac{4\bar{\omega}^2(1+\bar{n})^2+(1+2\bar{n})^2((1+\bar{n})^2+\bar{\Delta}^2)}{(1+2\bar{n})^2(4\bar{\omega}^2+(1+2\bar{n})^2+4\bar{\Delta}^2)}\nn\\
\rho_{14}&=&-\frac{\bar{\Delta}(2\bar{\omega}+i(2\bar{n}+1))}{(1+2\bar{n})(4\bar{\omega}^2+(1+2\bar{n})^2+4\bar{\Delta}^2)}
\eqa In the limit of zero temperature (i.e., $\bar{n}\rightarrow
0$), the density matrix elements in Eq.~(\ref{densitymatrixT})
reduce to the results in Eqs.~(\ref{p11a}) - (\ref{p41a}).  In the
limit of infinite temperature (i.e., $\bar{n}\rightarrow \infty$),
the density matrix becomes diagonal, with each diagonal element
equal to $1/4$, indicating, as expected {\cite{Fine}}, that all
entanglement vanishes.

The concurrence may be calculated for the finite temperature,
steady-state density matrix in Eq.~(\ref{densitymatrixT}) to obtain:

\bqa\label{CT}
C(\bar{\omega},\bar{\Delta},\bar{n})&=&2\frac{\sqrt{\bar{\Delta}^2(4\bar{\omega}^2+(1+2\bar{n})^2)}}{(1+2\bar{n})(4\bar{\Omega}^2+(1+2\bar{n})^2)}-\frac{1}{2}\nn\\&&+\frac{(4\bar{\omega}^2+(1+2\bar{n})^2)}{2[(1+2\bar{n})^2(4\bar{\Omega}^2+(1+2\bar{n})^2)]},
\eqa where all system parameters have been normalized by the
relaxation rate $\gamma$: $\bar{\Delta}=\Delta/\gamma$,
$\bar{\omega}=\omega/\gamma$, and $\bar{\Omega} =
\Omega/\gamma=\sqrt{\bar{\omega}^2+\bar{\Delta}^2}$ (cf.
Eq.~(\ref{omega})).  In the limit of zero temperature (i.e.,
$\bar{n}\rightarrow 0$), the concurrence in Eq.~(\ref{CT}) reduces
to that in Eq.~(\ref{steadycon}).  The behavior of this finite
temperature, steady state concurrence is shown in Fig.~{\ref{fig4}}
for the same two sets of system parameters considered in
Figs.~\ref{fig1} and~\ref{fig2} respectively.  One sees that both
curves decrease with increasing $\bar{n}$ until eventually the
concurrence vanishes at a finite value of $\bar{n}$, as expected
\cite{Fine}.  One sees also that the larger the value of the
interaction asymmetry parameter $\bar{\Delta}$, the larger the value
of the concurrence at any finite value of $\bar{n}$.  For any fixed
temperature (i.e., $\bar{n}$), as the effective magnetic field,
$\bar{\omega}$, increases, the concurrence takes a finite, non-zero
value.  In the limit $\bar{\omega}\rightarrow \infty$, one has that
$\bar{n}\rightarrow 0$ and $C\rightarrow
|\bar{\Delta}|/\bar{\omega}$.  This decrease with
$\bar{\omega}^{-1}$ as well as the fact that $C \ge 0$ only if
$\bar{\Delta} \not= 0$ is consistent with the results of
Ref.~\cite{gerard}.

\begin{figure}
\begin{center}
\includegraphics[width=8.0cm]{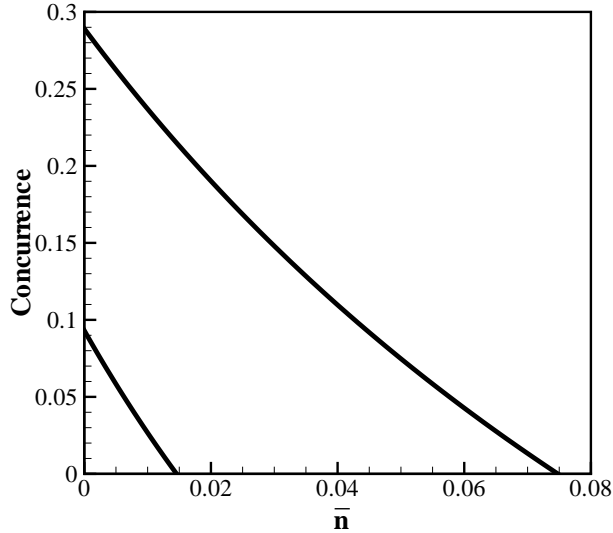}
\end{center}
\caption{Plots of the finite temperature, steady state concurrence
(cf. Eq.~(\ref{CT})) as a function of the average thermal excitation
function, $\bar{n}$, for the same two sets of system parameters as
in Figs.~\ref{fig1} and~\ref{fig2}.  The curve in the lower left
corner of the figure corresponds to the same system parameters as in
Fig.~\ref{fig1}; the curve close to the diagonal corresponds to the
same system parameters as in Fig.~\ref{fig2}. Note that at $\bar{n}
= 0$ both curves begin at the steady state values of the concurrence
shown in Figs.~\ref{fig1} and~\ref{fig2}.}\label{fig4}
\end{figure}

\section{\label{Discussion} Discussion and Conclusions}

Quantum coherence is a necessary requirement for the existence of
entanglement. One may define coherence existing in a single qubit as
``local coherence'' while coherence between two qubits $A$ and $B$
may be defined as ``global coherence'' \cite{Eberly}. How do local
and global coherence relate to entanglement? The answer for our
model system may be understood by considering the relation between a
general two-qubit density matrix, $\rho_{AB}$, and the reduced
density matrices, $\rho^A$ and $\rho^B$, for each of the two qubits,
where $\rho^A=tr_B(\rho^{AB})$ is obtained by tracing over the
degrees of freedom of qubit $B$, and similarly for $\rho^B$. In our
model, after doing partial traces, we find  that  in the steady
state the local coherence of each qubit is zero, i.e., $\rho^A$ and
$\rho^B$ are diagonal matrices. However, there exist global
coherence terms for our system (i.e., $\rho_{14}$ and $\rho_{41}$)
that are non-zero, indicating that global, not local, coherence is
responsible for this system's entanglement. For time $t>0$,
$\rho_{14}$ and $\rho_{41}$ for our system are always non-zero;
$\rho_{23}$ and $\rho_{32}$ may be non-zero for finite times, but
vanish in the steady state. Ref.~{\cite{Eberly}} considered a
decohering system of two entangled (but non-interacting) qubits.  In
that system,  both local and global coherence vanished in the
asymptotic time limit; however,  in some cases, the latter vanished
for finite times \cite{Eberly}.

An interesting question regarding the steady state of our model
system is whether or not it is a decoherence-free subspace
\cite{DFS}. Typically, a decoherence-free subspace is defined to be
one for which the decoherence terms in the system's master equation
vanish \cite{Whaley}.  For our system, this would mean that the
second and third terms on the right of Eq.~(\ref{Lindblad}) vanish
for the case of zero temperature or, for the case of finite
temperature, that all terms except for the first one on the right of
Eq.~(\ref{Lindblad2}) vanish.  However, in our system, the
decoherence terms in Eqs.~(\ref{Lindblad}) and~(\ref{Lindblad2}) do
not vanish; rather the sum of all terms on the right hand sides of
these two master equations vanish. This implies that population
relaxation and thermal decoherence (in the case of finite
temperature) are competing with the spin-spin interaction terms to
create a steady state level of entanglement, as measured by the
concurrence.

We note, finally, that a somewhat different model system studied by
S. Montangero, G. Benenti, and R. Fazio \cite{Montangero} has found
results for the pairwise concurrence that are somewhat similar to
those we find for our system.  Specifically, they have considered
the entanglement of a pair of spins within a qubit lattice in which
there is disorder in the spin-spin couplings.  They have identified
a regime in which the pairwise concurrence is stable against such
disorder in the couplings and has a value in the range of
0.2-0.3~\cite{Montangero}.  We note that the numerical maximum for
their ``saturation value'' of the concurrence is quite close to the
analytical maximum we have derived in this paper (cf.
Eq.~(\ref{Csteadybmax})). It is interesting to observe that our
analytic result for the maximum value of the steady state
concurrence is constant for a range of system parameters (cf.
Eqs.~(\ref{Deltamax})-(\ref{Csteadybmax})). Whether or not this
analytical maximum holds also for other systems, such as the
different one considered in~\cite{Montangero},  is an open question.

In summary, we have provided a detailed analytical and numerical
analysis of decoherence for an interacting  two-qubit model system
having a Hamiltonian identical in form to that for the well-known
two-qubit Heisenberg XY spin 1/2 system in the presence of an
(effective) external uniform magnetic field.  For $T=0$, we have
presented an analytic solution for the evolution of entanglement,
measured by concurrence, for the case that both qubits are initially
in their ground states; we have presented also numerical solutions
for two other typical initial states. We find that our system is
robust against decoherence: a steady state level of entanglement,
controllable by the values of the system parameters,  is always
reached for zero or finite, low temperatures.  For the $T=0$ case,
we have defined this steady state analytically and  obtained the
parameter values that maximize its entanglement.  For $T>0$, the
steady state level of entanglement is found to vanish at a finite
temperature. Since our model interaction Hamiltonian describes also
mesoscopic objects that interact via their spins (e.g., cf.
Ref.~\cite{Skomski}), it may be that a certain level of entanglement
is robust against decohering interactions with an environment even
for such objects.   As noted by Ghosh {\it et al.}~\cite{Ghosh},
even ``the slightest degree of entanglement can have profound
effects'' on the properties of mesoscopic spin systems.

We acknowledge stimulating discussions with Joseph H. Eberly, Hong
Gao, Andrei Y. Istomin, Murray Holland, Ting Yu, and Peter Zoller.
This work is supported in part by grants from the Nebraska Research
Initiative and the W. M. Keck Foundation.



\section*{References}

\begin{thebibliography}{10}


\bibitem{Bouwmeester} 2000 {\em The Physics of Quantum Information}, edited by Bouwmeester D, Ekert A, and Zeilinger A (Berlin:Springer)

\bibitem{Arnesen} Arnesen M C, Bose S, and Vedral V 2001 {\textit{Phys. Rev. Lett.}}
\textbf{87} 017901

\bibitem{Wang} Wang X 2001  {\textit{Phys. Rev.}} A \textbf{64} 012313

\bibitem{gerard}
Lagmago Kamta G and Starace A F 2002 {\textit{Phys. Rev. Lett.}} \textbf{88}
107901

\bibitem{Korepin} Korepin V E, Bogoliubov N M, and Izergin A G 1993 \emph{Quantum Inverse
Scattering Method and Correlation Functions} (Cambridge: Cambridge University
Press) pp. 63-79

\bibitem{Loss1} Loss D and DiVincenzo D P 1998 {\textit{Phys. Rev.}} A  \textbf{57} 120

\bibitem{Loss2} Burkard G, Loss D, and DiVincenzo D P 1999 {\textit{Phys. Rev.}} B \textbf{59} 2070 

\bibitem{Imamoglu} Imamo\={g}lu A \emph{et al.} 1999 {\textit{Phys. Rev. Lett.}}
\textbf{83} 4204

\bibitem{Ernst} Ernst R R, Bodenhausen G, and
Wokaun A 1988 {\em Principles of Nuclear Magnetic Resonance in One and
Two Dimensions} (Oxford: Clarendon Press)

\bibitem{Nielson} Nielson M A and Chuang I L 2000 {\em Quantum
Computation and Quantum Information} (Cambridge: Cambridge University Press).

\bibitem{SorensenMolmer} S{\o}rensen A and M{\o}lmer K 1999 {\textit{Phys. Rev. Lett.}} \textbf{83} 2274 

\bibitem{Duan} Duan L-M, Demler E, and Lukin M D 2003 {\textit{Phys. Rev. Lett.}} \textbf{91} 090402 

\bibitem{Shepelyansky} Georgeot B and Shepelyansky D L 2000 {\textit{Phys. Rev.}} E \textbf{62} 3504; 2000 \textit{ibid.}~\textbf{62} 6366

\bibitem{Makhlin} Makhlin Y, Sch{\"o}n G, and Shnirman A 2001 {\textit{Rev. Mod. Phys.}} \textbf{73} 357, Appendix A

\bibitem{Bloch} Bloch F 1946 {\textit{Phys. Rev.}} \textbf{70} 460

\bibitem{Zurek} Zurek W H 1991 {\textit{Phys. Today}} \textbf{44(10)} 36; 1993 {\textit{Prog. Theor. Phys.}} \textbf{89} 281; 2003 {\textit{Rev. Mod. Phys.}} \textbf{75} 715

\bibitem{Albrecht} Albrecht A 1992 {\textit{Phys. Rev.}} D \textbf{46} 5504

\bibitem{Das Sarma} de Sousa R and Das Sarma S 2003 {\textit{Phys. Rev.}} B \textbf{68} 115322; Hu X, de Sousa R, and Das Sarma S  2001 arXiv:cond-mat/0108339.

\bibitem{Dodd} Dodd P J and Halliwell J J 2004 {\textit{Phys. Rev.}} A \textbf{69} 052105

\bibitem{Lloyd} Viola L, Lloyd S, and Knill E 1999 {\textit{Phys. Rev. Lett.}} \textbf{83} 4888

\bibitem{Knight} Beige A, Braun D, Tregenna B, and Knight P L 2000 {\textit{Phys Rev. Lett.}} \textbf{85} 1762

\bibitem{Braun} Braun D, Haake F, and Strunz W T 2001 {\textit{Phys. Rev. Lett.}} \textbf{86} 2913

\bibitem{Eberly} Yu T and Eberly J H 2004 {\textit{Phys. Rev. Lett.}} \textbf{93} 140404

\bibitem{Bennett} Bennett C H, DiVincenzo D P, Smolin J A, and Wootters W K 1996 {\textit{Phys. Rev.}} A \textbf{54} 3824

\bibitem{Wootters} Hill S and Wootters W K 1997 {\textit{Phys. Rev. Lett.}} {\textbf{78}} 5022; Wootters W K 1998 {\textit{Phys. Rev. Lett.}} {\textbf {80}} 2245

\bibitem{Carmichael} Carmichael H J 1999  {\em Statistical Methods in Quantum Optics 1:Master Equations and Fokker-Planck Equations} (Berlin:Springer-Verlag) p. 16

\bibitem{Gardiner} Gardiner C W and Zoller P 2000 {\em Quantum Noise}, 2nd Ed. (Berlin: Springer-Verlag) p. 147

\bibitem{lindblad}
Lindblad G 1976 {\textit{Commun. Math.  Phys.}} {\bf 48} 199

\bibitem{Vedral} Vedral V, Plenio M B, Rippin M A, and Knight P L 1997 {\textit{Phys. Rev. Lett.}} \textbf{78} 2275

\bibitem{Horodecki} Horodecki M, Horodecki P, and Horodecki R 1998 {\textit{Phys. Rev. Lett.}} \textbf{80} 5239

\bibitem{Rains} Rains E M 1999 {\textit{Phys Rev.}} A {\textbf{60}}  173; \textbf{60} 179


\bibitem{Preskill} Preskill J
{\em Lecture Notes on Quantum Information and Quantum Computation}
at www.theory.caltech.edu/people/preskill/ph229

\bibitem{ZollerPC} Zoller P 2005 private communication

\bibitem{Fine} Fine B V, Mintert F, and Buchleitner A 2005 {\textit{Phys. Rev.}} B \textbf{71} 153105 

\bibitem{Carmichael2} See Ref.~\cite{Carmichael}, Eq.~(2.26)

\bibitem{Mintert} Mintert F, Carvalho A R R, Ku\'{s} M, and Buchleitner A 2005 quant-ph/0505162v1, Eq.~(110)

\bibitem{DFS} See, e.g., p.498 of Ref.~\cite{Nielson}.

\bibitem{Whaley} Lidar D A, Chuang I L, and Whaley K B 1998 {\textit{Phys. Rev. Lett.}} {\bf 81} 2594 

\bibitem{Montangero} Montangero S, Benenti G, and Fazio R 2003 {\textit{Phys. Rev. Lett.}} {\bf 91} 187901 

\bibitem{Skomski} Skomski R, Istomin A Y, Starace A F, and Sellmyer D J 2004 {\textit{Phys. Rev.}} A {\bf 70} 062307

\bibitem{Ghosh} Ghosh S, Rosenbaum T F, Aeppli G, and Coppersmith S N 2003 {\textit{Nature}} {\bf425} 48

\end{thebibliography}
\end{document}